%% file: TOP-16-020_temp.tex
\pdfoutput=1

\documentclass[11pt,twoside,a4paper,cmspaper,final,collab]{cms-tdr}
\def\svnVersion{449272P}\def\cmsCernNoTag{CERN-EP-2017-296}\def\cmsCernDate{\today}\def\cmsMessage{Published in Physics Letters B as \href{http://dx.doi.org/10.1016/j.physletb.2018.02.025}{\doi{10.1016/j.physletb.2018.02.025.}}}

\begin{document}\cmsNoteHeader{TOP-16-020}

\hyphenation{had-ron-i-za-tion}
\hyphenation{cal-or-i-me-ter}
\hyphenation{de-vices}
\RCS$Revision: 449218 $
\RCS$HeadURL: svn+ssh://svn.cern.ch/reps/tdr2/papers/TOP-16-020/trunk/TOP-16-020.tex $
\RCS$Id: TOP-16-020.tex 449218 2018-03-05 15:58:35Z mara $
\newlength\cmsFigWidth
\ifthenelse{\boolean{cms@external}}{\setlength\cmsFigWidth{0.85\columnwidth}}{\setlength\cmsFigWidth{0.4\textwidth}}
\ifthenelse{\boolean{cms@external}}{\providecommand{\cmsLeft}{top\xspace}}{\providecommand{\cmsLeft}{left\xspace}}
\ifthenelse{\boolean{cms@external}}{\providecommand{\cmsRight}{bottom\xspace}}{\providecommand{\cmsRight}{right\xspace}}
\providecommand{\NA}{\multicolumn{1}{c}{\ensuremath{\text{---}}}}

\newcommand{\mtw}{\ensuremath{m_\mathrm{T}^\PW}\xspace}
\newcommand{\tZq}{\ensuremath{\PQt\PZ \PQq}\xspace}
\newcommand{\tZllq}{\ensuremath{\PQt\ell^+\ell^- \PQq}\xspace}
\newcommand{\ttZ}{\ensuremath{\PQt\PAQt\PZ}\xspace}
\newcommand{\ttW}{\ensuremath{\PQt\PAQt\PW}\xspace}
\newcommand{\ttH}{\ensuremath{\PQt\PAQt\PH}\xspace}
\newcommand{\WZCR}{0bjet}
\newcommand{\tZqCR}{1bjet}
\newcommand{\ttZCR}{2bjets}
\newcommand{\Mell}{\ensuremath{m_{\ell^+\ell^-}}\xspace}
\newcommand{\eee}{\ensuremath{\Pe\Pe\Pe}\xspace}
\newcommand{\eemu}{\ensuremath{\Pe\Pe\Pgm}\xspace}
\newcommand{\mumue}{\ensuremath{\Pe\Pgm\Pgm}\xspace}
\newcommand{\mumumu}{\ensuremath{\Pgm\Pgm\Pgm}\xspace}

\cmsNoteHeader{TOP-16-020}

\title{Measurement of the associated production of a single top quark and a Z boson in pp collisions at $\sqrt{s} = 13$\TeV}

\date{\today}

\abstract{
A measurement is presented of the associated production of a single top quark and a Z boson.
The study uses data from proton-proton collisions at $\sqrt{s} = 13$\TeV
recorded by the CMS experiment, corresponding to an integrated luminosity of 35.9\fbinv.
Using final states with three leptons (electrons or muons), the  tZq production cross section is measured to be
$\sigma (\Pp\Pp\to\PQt\PZ\PQq\to\PW\PQb\ell^+\ell^-\PQq) =
123^{+33}_{-31}\stat^{+29}_{-23}\syst\unit{fb}$,  where $\ell$ stands for
electrons, muons, or $\tau$ leptons, with  observed and expected significances of 3.7 and 3.1 standard deviations, respectively.
}

\hypersetup{%
pdfauthor={CMS Collaboration},%
pdftitle={Measurement of the associated production of a single top quark and a Z boson in pp collisions at sqrt(s) = 13 TeV},%
pdfsubject={CMS},%
pdfkeywords={SM, single top, cross section, tZq}}

\maketitle

\section{Introduction}
\label{sec:Intro}

At the CERN LHC, single  top quark production
proceeds through three electroweak  interaction processes:
$t$-channel, $s$-channel, and associated $\PQt\PW$~production.
Cross sections for
single top quark production have been reported
by the  CDF and D0 Collaborations \cite{PhysRevLett.115.152003,PhysRevLett.112.231803},
as well as  by the ATLAS~\cite{ATLAStchan7TeV,ATLAStWchan8TeV,ATLASsChan,ATLASsingletop13TeV,ATLASsingletoptW13TeV} and CMS~\cite{CMStchan7TeV, CMStWchan8TeV,CMSschan, CMSsingletop13TeV} Collaborations.

\begin{figure*}[hbt]
\centering
\includegraphics[width=0.35\textwidth]{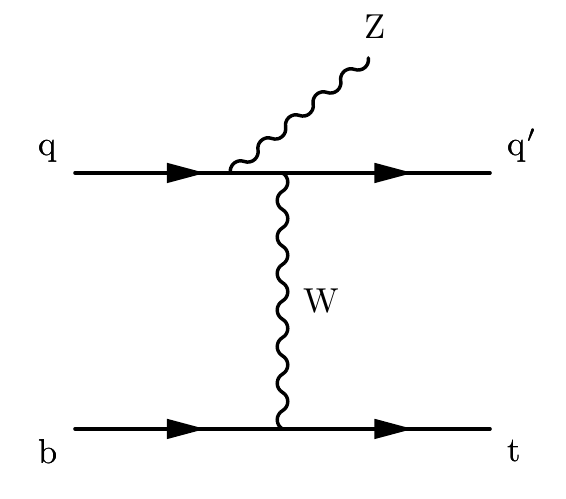}
\includegraphics[width=0.35\textwidth]{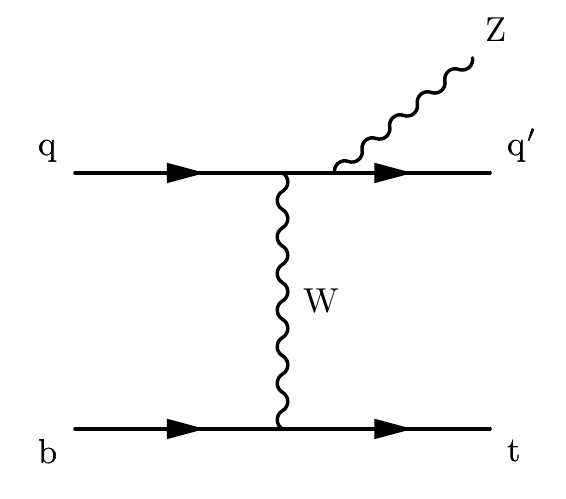}
\includegraphics[width=0.35\textwidth]{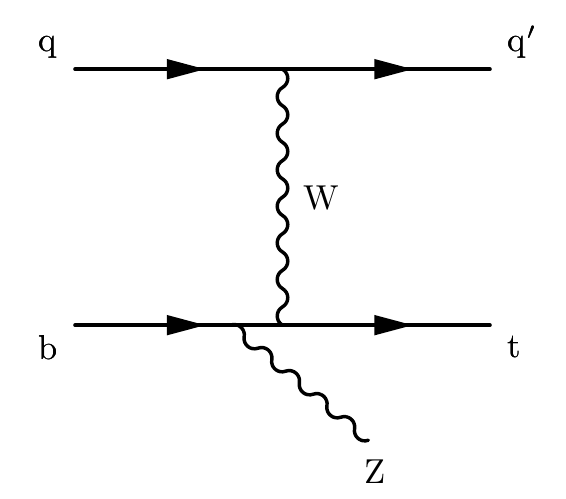}
\includegraphics[width=0.35\textwidth]{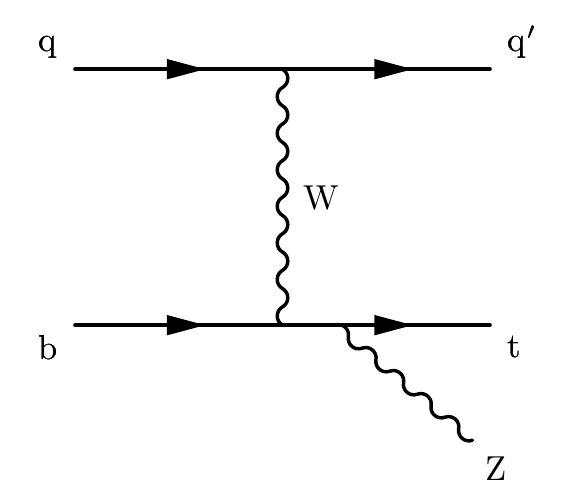}
\includegraphics[width=0.35\textwidth]{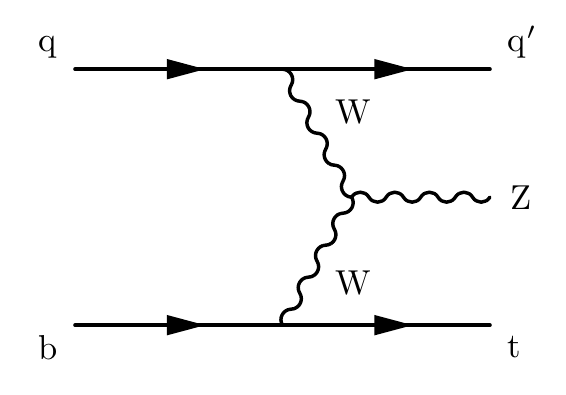}
\includegraphics[width=0.35\textwidth]{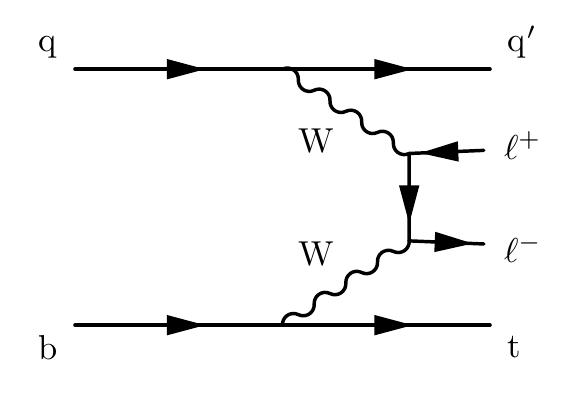}
\caption{Leading-order \tZq production diagrams. The lower right-hand diagram represents the non-resonant contribution to the \tZq process.}
\label{fig:feyn}
\end{figure*}

The high centre-of-mass proton-proton (\Pp\Pp) collision energy of 13\TeV at the LHC, together with large integrated luminosities,
allows the study of processes with very small cross sections that were not accessible at lower energies.
One example of such a process is the rare associated production of a single top quark with a \PZ boson.
This production mechanism, leading to a final state with a top quark, a \PZ boson, and an additional quark,
can probe the standard model (SM) in a unique way.
 The main leading-order (LO) diagrams that contribute to this  final state are shown in Fig.~\ref{fig:feyn}.
Although generically denoted in this Letter by \tZq, this process also includes a small contribution from non-resonant lepton pairs,
as shown in the lower right-hand diagram in Fig.~\ref{fig:feyn}.
The process  is sensitive to top quark couplings to the \PZ boson, as illustrated in the middle  right-hand diagram in Fig.~\ref{fig:feyn}, and also to the triple gauge-boson coupling $\PW\PW\PZ$, as illustrated in the lower left-hand diagram in Fig.~\ref{fig:feyn}.

The top quark couplings to the \PZ boson and the triple gauge-boson couplings are sensitive to new physical phenomena.
In particular, measurements of \tZq\ production are sensitive to processes beyond the SM that have similar experimental signatures,
such as flavour-changing
neutral currents (FCNC)
involving the direct coupling of the top quark to a \PZ boson and an up or charm quark,  at the top quark
production or decay \cite{AguilarSaavedra:2004wm,Agram:2013koa}.
Within the SM, FCNC processes are forbidden at LO and  suppressed at higher orders~\cite{GIM}.
Deviations from the expected SM \tZq production could therefore be indicative of beyond-SM FCNC processes.

The next-to-leading-order (NLO) cross section for $\tZq \to \PW \PQb  \ell^+ \ell^-  \PQq$,
considering only the leptonic decays of  \PZ bosons (to electrons, muons, or $\tau$ leptons, generically denoted by $\ell$),
is calculated for \Pp\Pp~collisions at a centre-of-mass energy of 13\TeV, using the Monte Carlo (MC) generator
 \MGvATNLO 2.2.2~\cite{amcatnlo}.
The calculation, which includes lepton pairs from off-shell \PZ bosons with invariant mass $\Mell>30$\GeV,
 uses the {NNPDF 3.0} set of parton distribution functions (PDFs)~\cite{nnpdf30} in the five-flavour scheme.
The result is
$\sigma^\mathrm{SM}(\tZllq)=94.2^{+1.9}_{-1.8}\,\text{(scale)}\pm 2.5\,\text{(PDF)}\unit{fb}$,
with the ``scale" and ``PDF" uncertainties estimated, respectively, by changing the  quantum chromodynamics (QCD)
renormalization and factorization scales by factors of 0.5 and 2,
and  by using the 68\% confidence level (CL) uncertainty on the {\sc NNPDF3.0} PDF set.
This cross section is used as the reference in this analysis.
Another calculation, including all \PZ~boson decays, gives a compatible cross section~when
the branching fraction to charged leptons is taken into account \cite{Campbell:2013yla}.
Previous searches for \tZq production  at 8\TeV by the CMS Collaboration~\cite{TOP-12-039} reported a signal with a significance of  2.4 standard deviations.
The ATLAS Collaboration recently reported a measurement of the \tZq~production cross section at 13\TeV~\cite{ATLAStZq} with
a significance of  4.2 standard deviations.

This Letter presents a search for \tZq production in $\Pp\Pp$ collisions at  $\sqrt{s}=13$\TeV, using  data  collected in 2016 by CMS,
corresponding to an integrated luminosity of 35.9\fbinv.
 The signature for \tZq production
 consists of a single top quark produced
in the $t$ channel, a \PZ boson, and an additional (``recoiling") jet
emitted at
pseudorapidity  $\abs{\eta} < 4.5$.
 The analysis uses events where the  \PZ boson decays to $\Pep\Pem$ or $\Pgmp\Pgmm$,
while the \PW~boson, produced in the decay of the top quark,
decays to a neutrino and an electron or a muon,
resulting in four possible final-state leptonic combinations:  \eee, \eemu, \mumue, and \mumumu.
 There will also be a small contribution from $\tau$ leptons decaying into electrons
or muons.
The final result reflects an extrapolation to include all decay modes involving $\tau$ leptons.
The measurement is based on a multivariate analysis, where boosted decision trees (BDTs)~\cite{BDT} are used
to enhance the signal-to-background separation. Several control regions are defined to better constrain the backgrounds,
each containing different contributions from signal and background processes.

\section{The CMS detector}
\label{sec:cms}

The central feature of the CMS apparatus is a superconducting solenoid of 6\unit{m} internal diameter, providing a magnetic field of 3.8\unit{T}. Within the solenoid volume are a silicon pixel and strip tracker, a lead tungstate crystal electromagnetic calorimeter (ECAL), and a brass and scintillator hadron calorimeter (HCAL), each composed of a barrel and two endcap sections. Forward calorimeters extend the pseudorapidity coverage provided by the barrel and endcap detectors. Muons are measured in gas-ionization detectors embedded in the steel flux-return yoke outside the solenoid.
The electron momentum is
evaluated by combining the energy measurement in the ECAL with the momentum measurement in the tracker. The momentum resolution for electrons with
transverse momentum, $\pt$,
around $45\GeV$ from $\PZ \to \Pe \Pe$ decays ranges from 1.7\% for nonshowering electrons in the barrel region to 4.5\% for showering electrons in the endcaps~\cite{Khachatryan:2015hwa}.
Muons are measured in the  range $\abs{\eta} < 2.4$, with detection planes made using three technologies: drift tubes, cathode strip chambers, and resistive plate chambers. Matching muons to tracks measured in the silicon tracker results in a relative transverse momentum  resolution for muons with $20 <\pt < 100\GeV$ of 1.3--2.0\% in the barrel and better than 6\% in the endcaps
\cite{Chatrchyan:2012xi}.
A more detailed description of the CMS detector, together with a definition of the coordinate system used and the relevant kinematic variables, can be found in Ref.~\cite{Chatrchyan:2008zzk}.

Events of interest are selected using a two-tiered trigger system~\cite{Khachatryan:2016bia}. The first level, composed of custom hardware processors, uses information from the calorimeters and muon detectors to select events at a rate of around 100\unit{kHz} within a time interval of less than 4\mus. The second level, known as the high-level trigger, consists of a farm of processors running a version of the full event reconstruction software optimised for fast processing, and reduces the event rate to less than 1\unit{kHz} before data storage.

\section{Online selection, reconstruction, and identification}
\label{sec:evtselec}

The data  are selected online using triggers that rely on the presence of either one, two, or three high-\pt leptons.
The lowest \pt~thresholds of the three-lepton triggers are 16, 12, and 8\GeV for  electrons, and 12, 10, and 5\GeV for muons;
the corresponding values for the dilepton triggers are 23 and 12\GeV for electrons, and 17 and 8\GeV for muons.
Triggers requiring the presence of at least one electron and at least one muon are also used.
For the baseline offline selection, a trigger efficiency of nearly 100\% is achieved by including single-lepton triggers with thresholds of 32 and 24\GeV for electrons and muons, respectively, in addition to the two- and three-lepton triggers.

The events are reconstructed using the particle-flow (PF) algorithm~\cite{CMS-PF}, which
reconstructs and identifies each individual particle with an optimised combination of information from the various elements of the CMS detector. The energy of the photons is directly obtained from the ECAL measurement, corrected for zero-suppression effects, while that of the electrons is determined from a combination of the electron momentum at the primary interaction vertex as determined by the tracker, the energy of the corresponding ECAL cluster, and the total energy of all bremsstrahlung photons spatially compatible with originating from the electron track. The energy of the muons is obtained from the curvature of the corresponding track. The energy of charged hadrons is determined from a combination of their momentum, measured in the tracker, and the matching ECAL and HCAL energy deposits, corrected for zero-suppression effects and for the response function of the calorimeters to hadronic showers. Finally, the energy of neutral hadrons is obtained from the corresponding ECAL and HCAL corrected energy deposits.
For each event,
jets are clustered from the PF  candidates using the
anti-\kt algorithm~\cite{Cacciari:2008gp,Cacciari:2011ma}, with a distance parameter of 0.4.
The reconstructed vertex with the largest value of summed physics-object $\pt^2$ is taken to be the primary $\Pp\Pp$ interaction vertex. The physics objects are the jets, clustered  with the tracks assigned to the vertex as inputs, and the associated missing transverse momentum, taken as the negative vector sum of the \pt of those jets.
All charged particles considered in this analysis are required to be compatible with originating from the primary interaction vertex.

The event selection relies on the concept of relative lepton isolation, reflected in the variable $I_\text{rel}$,
computed as the scalar sum of the \pt of all particles in a cone of radius $\Delta R=\sqrt{\smash[b]{(\Delta\eta)^2 + (\Delta\phi)^2}}$ around the lepton
(where $\phi$ is the azimuth), excluding the lepton, and divided by the lepton \pt.
The sum is then corrected for the
 neutral particles produced in extra $\Pp\Pp$~interactions within the same or
neighbouring LHC bunch crossings, referred to as pileup (PU) collisions.
For electrons, $\Delta R$ is set to 0.3, and the expected PU  within
 the isolation cone  is estimated from the median energy density
 per area of PU contamination.
Muon $I_\text{rel}$ uses $\Delta R=0.4$, and is corrected for the
average neutral PU energy inside the isolation cone, which has been
measured in multijet events to be one half of the energy coming from charged hadrons not associated with the primary vertex.
Electrons and muons are considered isolated if $I_\text{rel}$ is smaller than  0.06 and 0.15, respectively.

The data  with prompt leptons are contaminated by
genuine leptons from hadron decays (usually referred to as ``nonprompt leptons")
and by hadrons or jets misidentified as leptons (usually referred to as ``fake leptons").
In addition, nonprompt isolated electrons can arise from the conversion of photons.
For simplicity of notation, and given that these background sources
are evaluated with similar methods, based on control samples in data,  all such sources are referred to as
``not-prompt" leptons, or simply ``NPL", in this Letter.
 Data samples for evaluating the NPL background are built using objects
 reconstructed similarly to the prompt leptons, with two important differences.
First, while the prompt and not-prompt leptons
are identified using the same variables~\cite{Khachatryan:2015hwa,Chatrchyan:2012xi}, less stringent criteria are applied to the NPL sample.
Second, leptons are considered not-prompt only if they are not isolated, requiring
 not-prompt  electrons or muons to have   $I_\text{rel} >0.17$ or $>0.25$, respectively. In addition,
not-prompt electrons are required to have $I_\text{rel} <1$, removing a large fraction of
photons with $I_\text{rel}\approx 1$ and $\PZ+$jets events containing a low-$\pt$ jet misidentified as a high $I_\text{rel}$ electron.
Tight criteria to reject photon conversions~\cite{Khachatryan:2015hwa} are required for both prompt and not-prompt electrons.

The jet momentum is determined from the vectorial sum of all particle momenta in the jet, and is found in simulation studies to be within 5 to 10\% of the true momentum over the whole \pt spectrum and detector acceptance. Jet energy corrections are obtained
from simulation studies and confirmed with in~situ measurements
through the balance in dijet, multijet, photon+jet, and leptonic \PZ\!+jet events
~\cite{Khachatryan:2016kdb}. In the central region, the jet energy resolution is approximately
15\% at 10\GeV, 8\% at 100\GeV, and 4\% at 1\TeV.
Jets reconstructed  at angular distances  $\Delta R< 0.4$ from the selected leptons are not considered for further analysis.
As the region  $2.7 < \abs{\eta}< 3.0$ is particularly affected by noise, events with jets of $\pt<50\GeV$ in that region are
rejected.

Jets that originate from the hadronization of a \PQb~quark are identified (tagged)
using the combined secondary vertex (CSVv2) algorithm~\cite{Chatrchyan:2012jua,CMS-PAS-BTV-15-001}, which combines
various track-based variables with secondary-vertex variables to construct a discriminating
observable in the region $\abs{\eta} < 2.4$.
At the chosen operating point, the CSVv2 algorithm
has an efficiency of about 83\% to correctly tag \PQb~jets and a probability of 10\%
for mistagging gluons and light quarks, as estimated from
simulation studies of  multijet events.

The missing transverse momentum vector \ptvecmiss is defined as the projection onto the plane perpendicular to the beam axis of the negative vector sum of the momenta of all reconstructed PF objects in an event. Its magnitude is denoted by \ptmiss.
The transverse mass of the W boson is defined as
 $\mtw = \sqrt{\smash[b]{ 2 \pt \ptmiss [1-\cos(\Delta\phi)]} }$, where \pt is the transverse momentum of
the lepton produced in the \PW~boson decay, and
$\Delta\phi$ is the difference in azimuth between the direction of the
lepton and the direction of \ptvecmiss.

\section{Simulated events }
\label{sec:pdf}

Monte Carlo  simulated events are used  extensively in this measurement
 to
evaluate the detector resolution, the efficiencies and acceptance, and to estimate the contributions from
background processes that have topologies similar to the trilepton \tZq final state.

The \tZq signal samples  are generated at NLO precision using the \MGvATNLO 2.2.2 package~\cite{amcatnlo}.
The two main background processes,  $\PW\PZ$+jets and  top quark pair production in association with vector bosons
(\ttZ and \ttW), are also simulated  with
the same event generator,
with up to one additional hadronic jet at NLO.
Other minor backgrounds are  $\PZ\PZ$ and \ttH~production, for which we use the NLO generators
\MGvATNLO  and \POWHEG v2.0~\cite{Nason:2004rx,Frixione:2007vw,Alioli:2010xd,Re:2010bp,Alioli:2009je,Melia:2011tj},
respectively,  and  $\PQt\PW\PZ$~production, generated
at LO accuracy using  \MGvATNLO.
The  PDF set {{NNPDF 3.0}} is used in all generators.
The simulated samples are interfaced to \PYTHIA 8.205~\cite{pythia8}
with the {CUETP8M1} tune~\cite{mctune} for the parton shower and hadronization.
The detector response is simulated using the \GEANTfour package~\cite{Agostinelli:2002hh}.

The events are simulated in final states that include decays to electrons, muons, and $\tau$ leptons.
A top quark mass of 172.5\GeV is assumed.
Multiple minimum-bias events generated with \PYTHIA are added to each simulated event to mimic
the presence of PU,
with weights that reproduce the measured distribution of the number of PU vertices.

The event samples are normalized to their expected cross sections, obtained from NLO calculations for all processes, except for $\PQt\PW\PZ$, which is estimated at LO accuracy.

Correction factors that depend on the $\pt$ and $\eta$ of the jets and leptons are applied to the samples,
 so that the resolutions, energy scales, and efficiencies measured in data are well reproduced by the simulation.
The  corrections include an extra  smearing of the jet energy, which has a better resolution in the simulation than
found in  data, and scale factors that account for different efficiencies in  lepton identification  and reconstruction.
The shape of the distribution in the CSVv2 discriminant is one of the variables used in the multivariate analysis to extract the
signal.
The simulated shape has been corrected~\cite{CMS-PAS-BTV-15-001,Chatrchyan:2012jua} to assure that the b tagging efficiency and purity
variables reproduce those found in data.

One of the most abundant background sources in the three-lepton final state
arises from events with at least one NPL.
Unlike all
other backgrounds, which are  modelled by MC simulation, the samples
used  to estimate the NPL background contribution
are obtained from the data, as described in Section~\ref{sec:NPL}.

\section{Event selection: signal and background control regions }
\label{sec:CR}

The event selection makes use of \tZq event candidates where
$\PQt\to \PW \PQb$, $\PW \to l \nu$, and $\PZ\to l'^{+} l'^{-}$,
\begin{linenomath}
\begin{equation*}
\label{eq:tzq}
\tZq \to(\PQt\to \PQb  l \nu)\ (\PZ\to l'^{+} l'^{-}) \PQq,
\end{equation*}
\end{linenomath}
where $l$ and $l'$ are either electrons or muons, coming  from the \PW~or \PZ~boson decay, respectively,
as opposed to generic leptons (including  $\tau$ leptons), which have been denoted by $\ell$.
As stated in Section 1, $l$ includes a small contribution from $\PW \to \tau \nu$ decays, with the
subsequent decay of the $\tau$ into electrons or muons. The final result will be given for all decay modes
to a generic lepton, $\ell$.
In single top quark production, the associated recoil jet usually follows the direction of the incoming proton,
so it is detected in the very forward regions of the detector. For this reason, we select jets in the extended pseudorapidity range  $\abs{\eta}<4.5$.
Given the tracker acceptance, b-tagged jets are  confined to the $\abs{\eta}<2.4$ range.
All jets, both tagged and untagged, are required to have $\pt>30$\GeV.

The baseline selection for the analysis consists
in  exactly three leptons, two of which have the same flavour, are oppositely charged, and
have an invariant mass compatible with the \PZ boson mass  within 15 \GeV.
Electrons and muons are  required to have $\pt>25$\GeV, and to be measured within $\abs{\eta}<2.5$ and 2.4, respectively.
To reduce backgrounds from four or more leptons in the final state, \eg from $\PZ\PZ$, \ttZ, and \ttH,
events containing additional leptons with $\pt>10$\GeV  and passing looser identification criteria are removed from the analysis.

Several other SM processes, some of which have much larger cross sections than
 expected for \tZq, contain  three reconstructed leptons
in the final state. Out of these, the most important are the $\PW\PZ$+jets, the \ttZ, and
those contributing to the NPL background.
For the first two, the three-lepton topology is identical to  \tZq:
two oppositely charged leptons of same flavour decaying from the \PZ boson, and a third
high-\pt, isolated lepton.
The \ttZ~production for the four-lepton final state has a smaller cross section than that for the
three-lepton final state, and is also suppressed by the already mentioned veto on events with four or more leptons.
Although the misidentification rate per lepton, especially for muons, is small, the
cross sections of the processes producing the NPL background
(dominated by Drell--Yan production in
association with jets, DY+jets, and \ttbar~production)
 are orders of
magnitude larger than the expected \tZq cross section, making NPL
 one of the most important
backgrounds to the three-lepton final state.

For the \tZq final state, two jets are expected, one of which arises from a b quark. In the
\ttZ~three-lepton final state, two b jets are expected. However, given the inefficiencies of the b tagging algorithm,
one of the two b jets may be untagged, leading to a final state identical to  the signal.
Likewise, one of the b jets  produced by gluon splitting in the $\PW\PZ$+jets final states may be tagged, or,
most frequently, light-flavour jets from $\PW\PZ$+jets production can be mistagged as b jets,
again resulting in a topology identical to the signal.

To reduce the impact of the background-related uncertainties on the measurement of
the \tZq yield, we proceed as follows. The
 baseline three-lepton selection is subdivided into three regions of interest, one enriched in \tZq events,
another selected to contain mostly \ttZ~events, and a third containing mostly $\PW\PZ$+jets and NPL background events.
The final analysis performs a simultaneous fit to these three regions, so that the signal cross section is determined
and the normalizations of the main backgrounds are better constrained.

The three regions are defined according to their jet and b-tagged jet multiplicities, as follows:
\begin{enumerate}
\item {\tZqCR~(signal region):} defined to select events from \tZq production
with one b jet and one recoiling jet. Events with a third jet are also included, to
cover cases where an additional jet is produced by radiation.
\item {\ttZCR~control region (\ttZ~enriched):}  defined by requiring at least two jets,
with at least two of them b tagged, enhancing thereby the yield in  \ttZ~events.
\item{\WZCR~control region ($\PW\PZ$+jets~enriched):} defined by at least one jet, but no  b-tagged jets,
selected as  most likely originating from a \PW\PZ process. Since the majority of DY+jets events also
do not contain b jets, this region is also rich in NPL background events.
\end{enumerate}

\section{Shape-based analysis}
\label{sec:bdt}

The \tZq cross section is extracted from a  binned maximum-likelihood fit
to the distributions in
the BDT discriminators (to be defined later) in the \ttZCR~and \tZqCR~regions,
and to the  \mtw\ distribution in the \WZCR~region.
Normalized distributions (templates) are constructed using these variables in their respective regions, for each of the four
final states (\eee, \eemu, \mumue, and \mumumu), adding up to 12 distributions that are
simultaneously fitted.

\subsection{Input normalization of the SM predictions}

The input (pre-fit) normalizations of the simulated backgrounds reflect their corresponding theoretical cross sections.
The contributions from \PW\PZ{}+\PQb, \PW\PZ{}+\PQc,~and \PW\PZ{}+light-flavour jets
in the \PW\PZ{}+jets MC events are separated using generator-level information, and considered as independent backgrounds
in all steps of the analysis. This provides a better modelling
of the heavy-flavour content of the \PW\PZ{}+jets sample,
and avoids relying on the flavour content of the MC simulation.

\subsection{The NPL background}
\label{sec:NPL}

The templates for the NPL background are
based on data.
The origin of not-prompt leptons depends on the lepton flavour.
For muons, the dominant source is the semileptonic decay of heavy-flavour hadrons.
In the case of electrons, the dominant sources are photon conversions and light hadrons that are misreconstructed as electrons.
The not-prompt electrons and muons are therefore treated as separate background sources.

The background events containing not-prompt leptons originate from, in order of importance, DY+jets processes,
 \ttbar events containing two leptons, and
 $\PW\PW$~and $\PQt\PW$~processes. Each of these background sources  contain two prompt and one not-prompt leptons.
Given the low probability that an NPL is identified as a prompt lepton,
the contribution from events with more than one NPL is negligible.
Not-prompt electron (muon) templates are obtained from events
containing exactly one not-prompt electron (muon), identified as
described in Section 3, and two prompt leptons (either electrons or
muons). In the NPL sample, the not-prompt leptons can be
associated either with the top quark  or with the Z boson
candidates.

The samples  used to obtain the NPL background templates are
quite copious, typically having two orders of magnitude more events than the signal
sample obtained with the baseline selection.
While the shapes of the distributions used in the multivariate analysis are provided by
 templates,
their normalizations are
determined through a two-step procedure. In the first step, the \mtw
distribution in the \WZCR~control region provides the
normalization of all NPL components, independently in the
four channels. This  fixes the relative NPL normalization
of the templates in the four channels.
In a subsequent step, the not-prompt electron and muon yields are treated as
 free and independent
parameters,  in a simultaneous fit of the
\WZCR/\tZqCR/\ttZCR~regions. This second step represents the final fit
used to provide the results reported in this Letter.

The use of the \WZCR~region to provide the relative NPL
yields in the four channels is justified by the dominance of the DY
process as source of NPL background events in all three b tagging
regions.
To check the validity of the procedure, an independent analysis
is performed where the weight of the DY background
relative to \ttbar~production is suppressed by means of mild requirements on
\ptmiss and \mtw. In this cross-check analysis, the relative
normalizations of the not-prompt electron and muon backgrounds are left free in the four channels, and
the results are obtained
in a single common fit. This alternative procedure gives similar final
results.

\subsection{Multivariate analysis}

The signal extraction relies on a simultaneous fit to the data in the three regions defined in Section \ref{sec:CR}, to better constrain the backgrounds in
the signal region.

Two multivariate discriminators,
 based on observables from the \tZqCR~and \ttZCR~regions,
 are used  to enhance the separation between signal and background processes.
The discriminators are based on the BDT algorithm~\cite{BDT} implemented in the
toolkit for multivariate analysis TMVA~\cite{Hocker:2007ht}.
The BDT is trained using the simulated samples described in Section \ref{sec:pdf}.

Several observables serve as input variables for the BDT.
These include the reconstructed top quark mass and distributions of variables reflecting
the kinematics and the angles of the recoiling jet, of the top quark, and of the \PZ boson,
as well as those of their decay products.
Once the two oppositely charged leptons of same flavour are identified as \PZ boson decay products, the additional lepton is assumed to arise from the decay $\PW\to l \nu$.
The longitudinal component of the neutrino momentum is calculated using the \PW~mass constraint for the $l +\nu$ system, and assuming the event \ptmiss~to be equal to the transverse momentum of the neutrino.
The reconstructed \PW~boson candidate is then associated to a b-jet candidate for the $\PQt\to \PW \PQb$ hypothesis.
The b-jet candidate is the tagged jet. If two solutions are found
for the longitudinal component of the neutrino momentum, or if more than one jet is tagged (in the \ttZCR~region), the solution giving the $\PW \PQb$ candidate invariant mass closest to that of the top quark is taken.
The remaining jet with the largest \pt~is taken as the recoiling jet.
The information related to b tagging is also used through the distributions of the CSVv2 discriminant~\cite{CMS-PAS-BTV-15-001,Chatrchyan:2012jua}
and the  b-tagged jet multiplicity.

Variables computed using the matrix element method (MEM)~\cite{Abazov:2004cs} are also included in the multivariate analysis.
A weight $w_{i,\alpha}$ is computed for each event $i$ and hypothesis $\alpha$ (where $\alpha$ is either signal, \ttZ, or \PW\PZ\!+jets) as
\begin{linenomath}
\ifthenelse{\boolean{cms@external}}{
\begin{multline*}
        w_{i,\alpha}(\Phi') = \frac{1}{\sigma_{\alpha}} \int \rd\Phi_{\alpha} \, \delta^4 \Big( p_1^{ \mu}+p_2^{ \mu} - \sum_{k>1}p_k^{ \mu} \Big) \\ \times\frac{  f(x_1,\mu_\mathrm{F}) f(x_2,\mu_\mathrm{F}) }{ x_1 x_2 s } \,\Big| \mathcal{M}_{\alpha} ( p_k^{ \mu} ) \Big|^2 \,W(\Phi'| \Phi_{\alpha}),
\end{multline*}
}{
\begin{equation*}
        w_{i,\alpha}(\Phi') = \frac{1}{\sigma_{\alpha}} \int \rd\Phi_{\alpha} \, \delta^4 \Big( p_1^{ \mu}+p_2^{ \mu} - \sum_{k>1}p_k^{ \mu} \Big) \, \frac{  f(x_1,\mu_\mathrm{F}) f(x_2,\mu_\mathrm{F}) }{ x_1 x_2 s } \,\Big| \mathcal{M}_{\alpha} ( p_k^{ \mu} ) \Big|^2 \,W(\Phi'| \Phi_{\alpha}),
\end{equation*}
}
\end{linenomath}
where: $\sigma_{\alpha}$ is the cross section; $\Phi'$ are the 4-momenta of the reconstructed particles;
$\rd\Phi_{\alpha}$ is the element of phase space corresponding to
parton-level variables
 with momentum conservation enforced \cite{Artoisenet:2010cn}; $f(x,\mu_\mathrm{F})$ are the PDFs,
where $\mu_\mathrm{F}$ is the QCD factorization scale, computed using the {NNPDF2.3LO} set~\cite{nnpdf}; $\abs{\mathcal{M}_{\alpha}}^2$ is the squared matrix element, computed with \MGvATNLO standalone~\cite{amcatnlo} at LO accuracy, in a narrow-width approximation for
the top quarks; and $W$ are the transfer functions for jet energy and \ptmiss,
relating parton-level variables to reconstructed quantities, evaluated from simulation studies and normalized to unity.

For all  three processes, the mass of the $\PW$ boson arising from the top quark
decay follows a Breit--Wigner distribution, as specified in the matrix element.
The virtual \PZ boson in the \ttZ~hypothesis  also follows a Breit--Wigner form, and
interference with $\Pgg^{*}$ is included in the matrix element.
The matrix element provided at LO in  \MGvATNLO does not contain additional jets
that are present in the data.
To evaluate the matrix element at LO, the momentum of the \tZq~system must have a null transverse component. The \tZq~momentum is computed as the sum of the momenta of all particles from the \tZq~decay. An inverse boost corresponding to the opposite of the \tZq~\pt is applied to all final state particles,
correcting thereby any recoiling jets not present in the LO matrix element.

In computing the MEM weights, jets with the highest CSVv2 discriminant values are assigned to the b quarks from top decays.
Among the remaining jets, up to two jets  with the highest $\abs{\eta}$ (signal hypothesis), with invariant mass closest to the
$\PW$~boson mass (for the \ttZ~ hypothesis),
or with the highest \pt (for the \PW\PZ\!+jets hypothesis), are assigned to the quarks at parton level.
Jets in the \tZqCR~region may not be matched to all parton-level quarks needed in the \ttZ~hypothesis (two b-quarks and two not-b quarks). In such cases, the
\ttZ~weight can still be computed by leaving the phase space of the missing jets unconstrained in the integral.

\begin{figure}[htb]
  \centering
\includegraphics[width=0.48\textwidth]{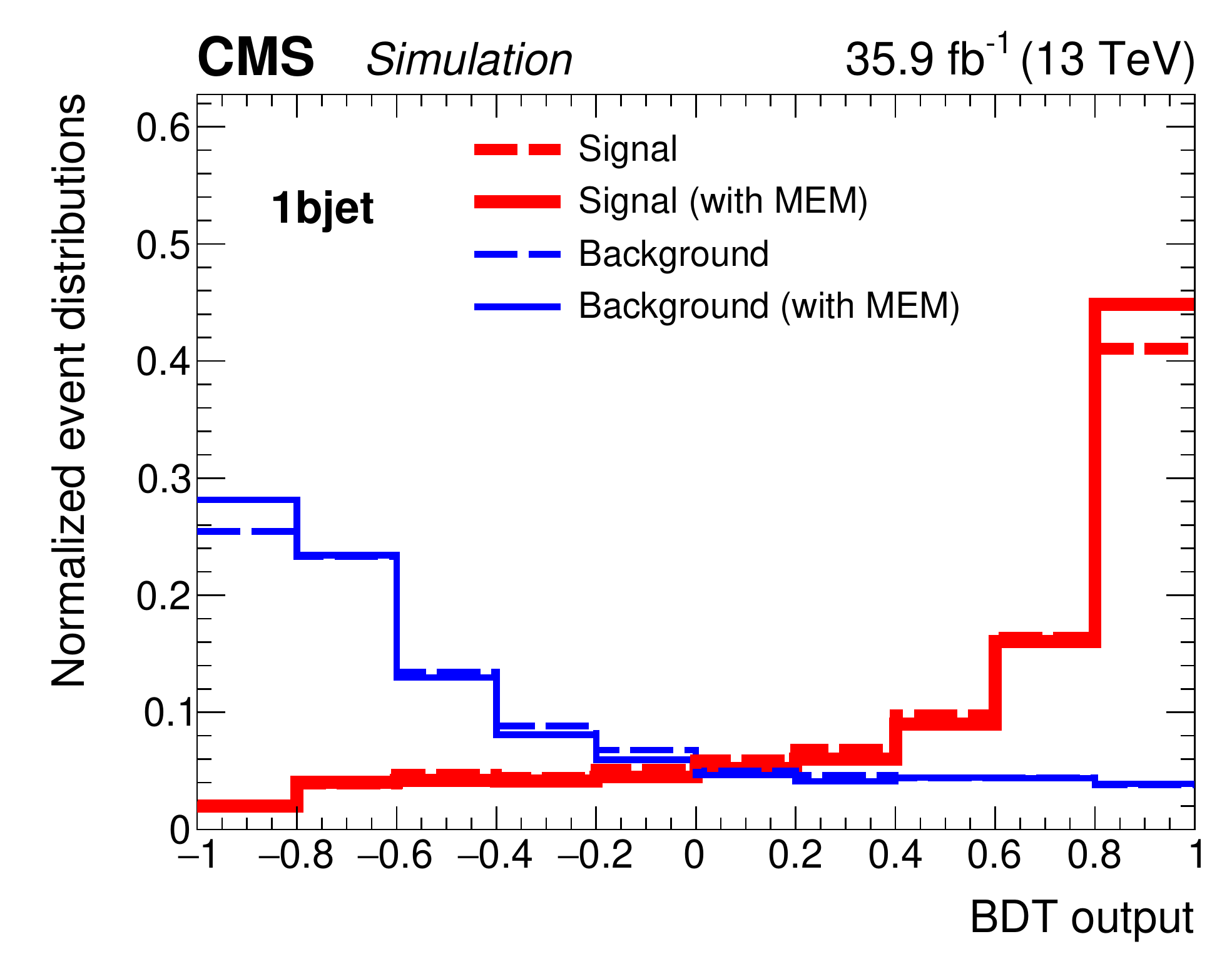}
\includegraphics[width=0.48\textwidth]{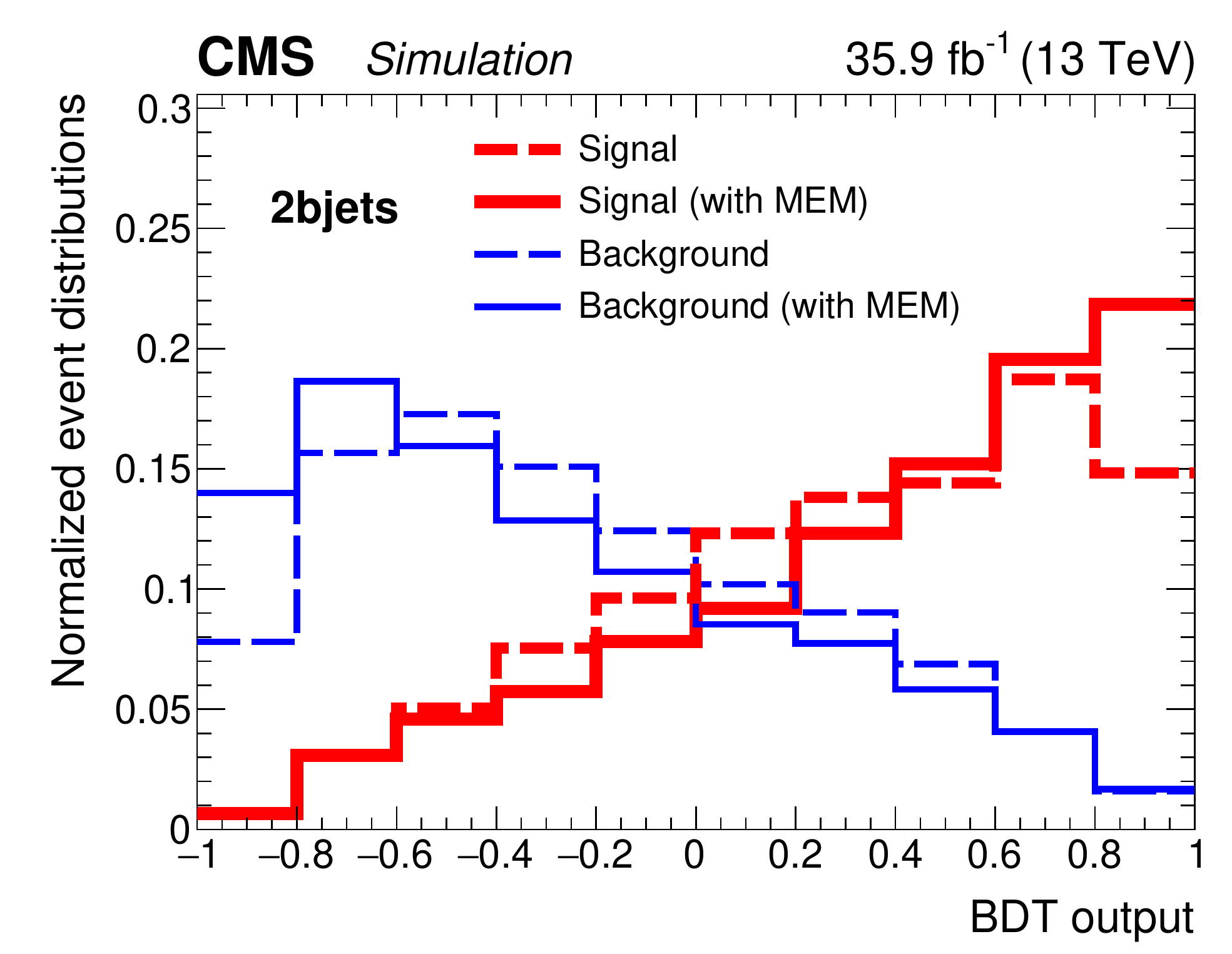}
    \caption{Normalized distributions of the BDT output for signal (thick lines) and backgrounds (thin lines)
from simulation for the \tZqCR~(\cmsLeft) and \ttZCR~(\cmsRight) regions. The discriminators including
and excluding MEM variables
in the BDT training are shown, respectively, as solid and dashed lines.
Contributions from the four considered channels are included in the signals and backgrounds. }
    \label{fig:membdtshape}
\end{figure}

The final weight for each hypothesis $\alpha$ is taken as the average of the weights computed for each lepton and jet permutation.
The MEM weights are combined in likelihood ratios of signal
to the combination of \ttZ~and \PW\PZ\!+jets in the \tZqCR~region and  signal to \ttZ~in the \ttZCR~region. These ratios
 are included  as input variables to the BDT.
In addition, the maximum value of the function being integrated is also included, corresponding to the MEM score associated to the most probable kinematic configuration.
Eight variables were tested and five were retained for the training; the other three were excluded because they were highly correlated
with other variables or had a negligible discriminant power.
The normalized BDT discriminators for signal and backgrounds in the \tZqCR~and \ttZCR~regions are
 shown in Fig.~\ref{fig:membdtshape} for BDT trainings with and without MEM variables.
Including the
MEM variables  improves the expected significance by about 20\%.

\begin{figure*}[htb]
        \centering
\includegraphics[width=\textwidth]{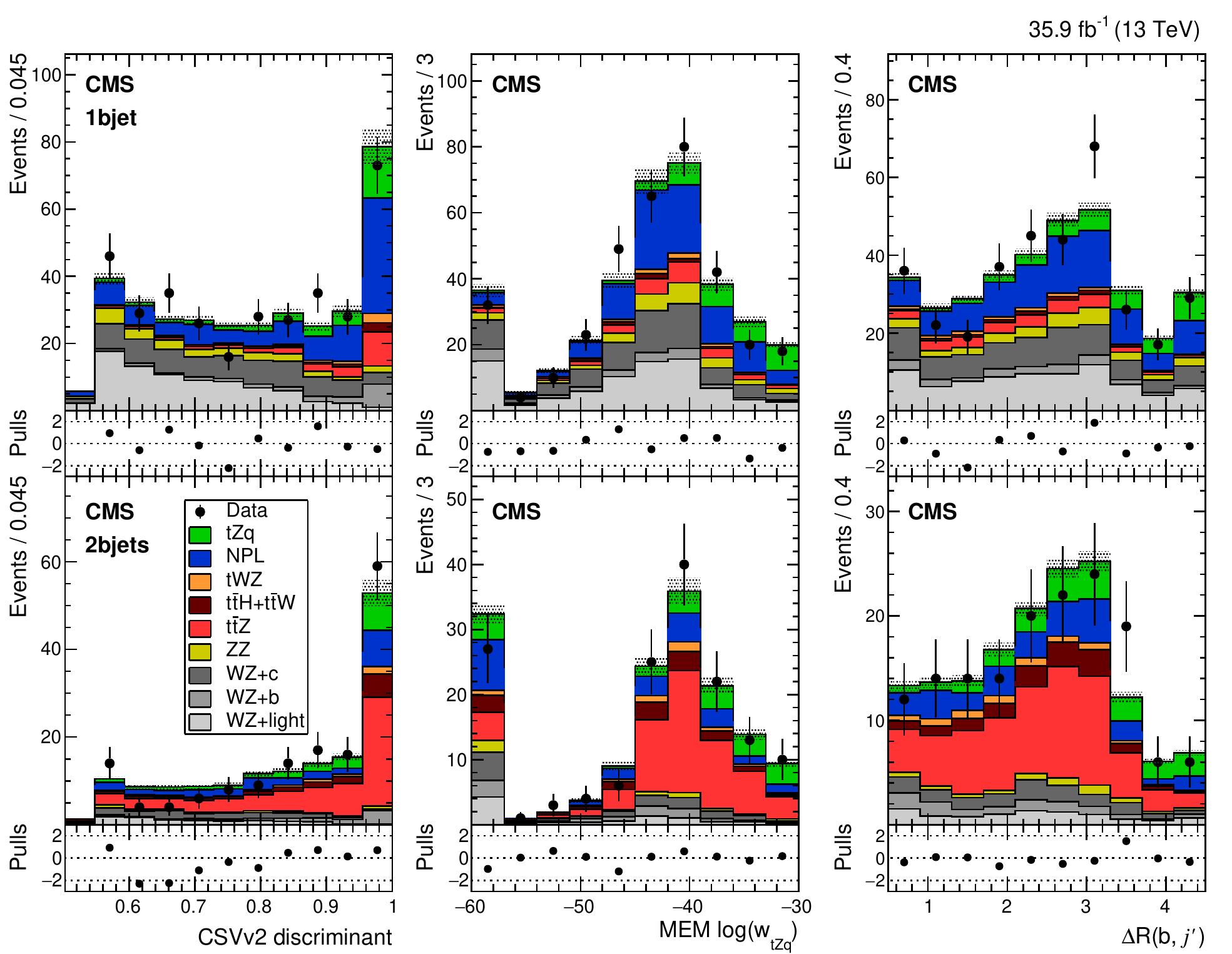}
    \caption{Data-to-prediction comparisons in the \tZqCR~region (signal-enriched, upper row) and
in the \ttZCR~region (lower row) for
 the  largest CSVv2 discriminant value among all selected jets (left), the logarithm of the MEM score associated to the most probable \tZq kinematic configuration (centre), and the
$\Delta R$ separation between the b quark and the recoiling jet (right).
The distributions include events from all final states. Underflows and overflows are shown in the first and last bins,
respectively.
The predictions correspond to the normalizations obtained after the fit described in Section \ref{sec:results3l}.
The hatched bands include the total uncertainty on the background and signal contributions. The pulls in the distributions are shown in the bottom panels. } 
    \label{fig:signal}
\end{figure*}

\begin{figure*}
        \centering
\includegraphics[width=\textwidth]{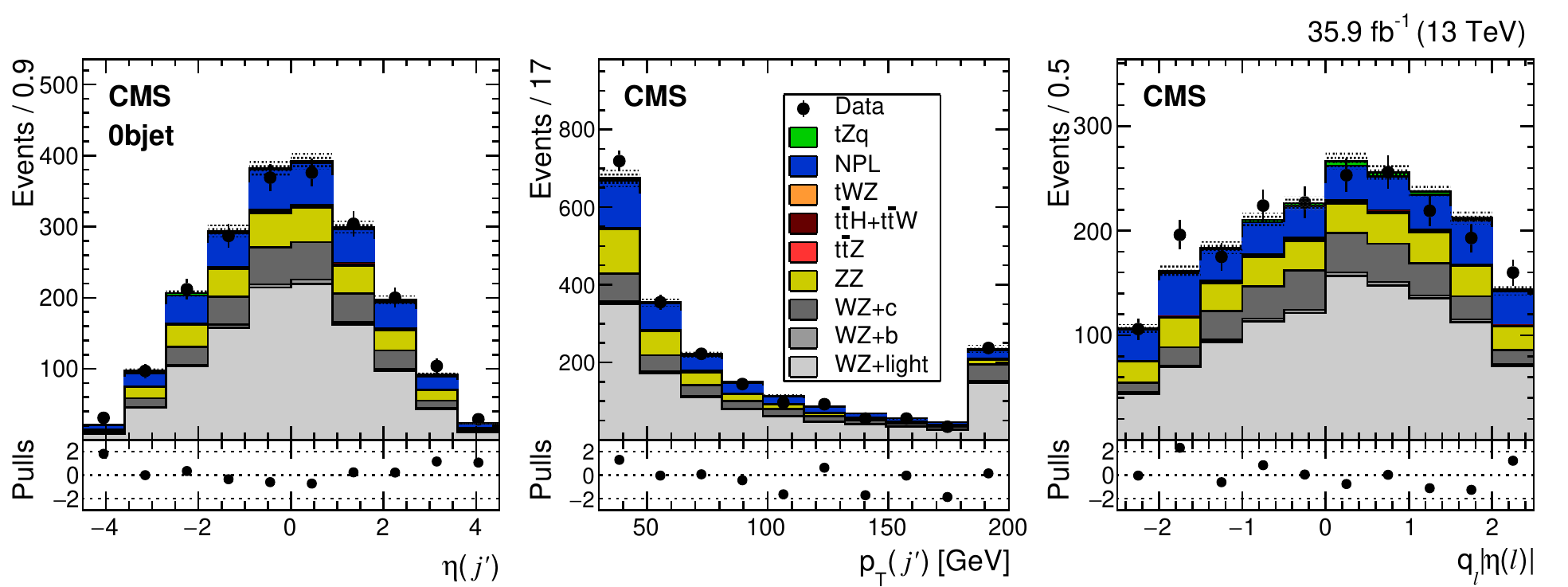}
    \caption{Data-to-prediction comparisons in the \WZCR~region for the
$\eta$  (left) and \pt (centre) distributions of the recoiling jet, and for the asymmetry
of the top quark decay lepton (right).
More details are given in the caption of Fig. \ref{fig:signal}.
    \label{fig:mtwWZ} }
\end{figure*}

The predictions for some of the most discriminating variables in the BDT for the \tZqCR~and \ttZCR~regions are compared to data
in Fig.~\ref{fig:signal}.
These variables are
the largest CSVv2 discriminant value among all selected jets, the logarithm of the MEM score associated to the
most probable \tZq kinematic configuration,
and the $\Delta R$ separation between the jet identified as a b quark and the recoiling jet.
Figure \ref{fig:mtwWZ} shows, for events in the \WZCR~region, the $\eta$ and
\pt distributions of the recoiling jet, $\eta({j'})$ and $\pt({j'})$, and
the  asymmetry of the top quark decay lepton, defined as the product of its charge and pseudorapidity, ${\PQq}_{l}|\eta(l)|$.
The distributions in Figs.~\ref{fig:signal} and~\ref{fig:mtwWZ} are shown combined for the  four channels: eee, ee$\mu$, e$\mu\mu$, and $\mu\mu\mu$.
The quadratic sum of the systematic and statistical
uncertainties on the predictions is shown as a hatched band.
The pulls of the distributions, defined in each bin as the difference between data and prediction, divided by the quadratic sum of total uncertainties
in the predictions (systematic and statistical) and the data (statistical), are shown at the bottom of the plots.

The complete list of variables used in the two BDTs is given in Appendix \ref{appBDT}.

\section{Systematic uncertainties}
\label{sec:syst}

Different sources of systematic uncertainty  can affect  the number of events passing the selections,
or the shape of the distributions used in the multivariate analysis.

The sources of systematic uncertainty considered correspond to:
\begin{itemize}

\item {Luminosity:} An uncertainty of 2.5\% on the sample integrated luminosity~\cite{LUM-15-001}
 is propagated  as a normalization-only uncertainty for the total predicted yields.
\item {Correction factors applied to the signal and simulated backgrounds:}
\begin{itemize}
\item {Pileup:} The number of simulated pileup events is corrected to match the measured number of events in
data. The uncertainty on the total inelastic cross section is taken as 4.6\%, and considered only in the shapes
of the distributions.
\item {Trigger:}  The trigger efficiency is estimated to be near 100\% both in data and in simulation.
Variations  in normalization of $\pm$1\% ($\pm$2\%) are applied to the predicted
yields in the \mumumu~and \eemu~(\mumue~and \eee) channels to account for residual differences
in trigger efficiency between data and simulation.
\item {Lepton selection:} The  factors used to correct the simulated distributions for lepton isolation and
identification efficiencies are varied by their uncertainty, affecting both the shapes and the normalizations.
\item  {Jet energy scale and resolution:}  The jet energy scale and  resolution
corrections are both varied by their  uncertainty.
The observed change is propagated to  all related kinematic quantities, in particular  \ptvecmiss.
These uncertainties affect both the shape and the normalization of the simulated distributions.
\item  {b tagging:}  The scale factors related to b tagging and mistagging efficiencies  are
varied by one standard deviation. Eight independent changes are considered,
including two types of statistical uncertainties
on the \PQb{}-, \PQc{}-, and light-flavour components of the MC event samples,
light-flavour  contamination of the b tagging scale factors,
and b quark contamination in the mistag scale factors. There is one ``nuisance"
parameter for each variation. Both shape
and normalization are affected.
\end{itemize}

\item {The  normalization of the simulated backgrounds:}
 The input normalizations of all simulated background distributions are assumed to have a relative uncertainty of 30\%.
 This  reflects the theoretical uncertainties on the corresponding cross sections, scaled up by a factor of two or more, to account for possible limitations in the simulations
in the phase space of the analysis.
  \item {The NPL background estimation:}  The shape-related uncertainties on the backgrounds involving
not-prompt leptons, determined with control samples in data, are estimated by
varying the isolation criteria used to determine the NPL sample.
The shape variations of not-prompt muons and electrons involve  different nuisance parameters.

\item {The scale and PDF uncertainties for
 simulated signal (\tZq) and background processes:} These uncertainties affect the shape of the signal as well as
 the shape and  normalization of the simulated background distributions,
except for $\PQt\PW\PZ$ events, for which only normalization uncertainties from scales and PDF are considered.

\begin{itemize}
\item  The renormalization and factorization scales, at the matrix element level, are set to an identical value,
which depends on the event generator and on the simulated processes. In particular, the scales for the simulated signal
are set to $\sum \sqrt{(m^2+\pt^2)}/2$, where the sum runs over all particles in the final state. The scales
are varied up and down by a factor of 2.
\item  The renormalization and factorization scales at the parton shower level, identical to the matrix element scales,
are also varied by  factors of 0.5 and 2; this uncertainty
is only evaluated for the signal sample.
\item The PDF uncertainties are evaluated by the root-mean-square of the results from  100 variations of the NNPDF set.
\end{itemize}
\end{itemize}

\section{Results}
\label{sec:results3l}

The tool used for this statistical analysis~\cite{higgscombine} is based on the {RooStats} framework~\cite{RooStats}.
The analysis is performed beginning with a binned likelihood function
\begin{linenomath}
\ifthenelse{\boolean{cms@external}}{
\begin{multline*}
{\mathcal{L}}(\text{data}|\mu,\theta) = \\
\sum_i\frac{\left[\mu s_i(\theta) + b_i(\theta) + \alpha_{\Pe} B_{i}^{\Pe}(\theta) +  \alpha_{\mu} B_{i}^{\mu}(\theta) \right]^{N_i}}{N_i !}\\
\times \re^{-\mu s_i(\theta) - b_i(\theta) - \alpha_{\Pe} B_{i}^{\Pe}(\theta) - \alpha_{\mu} B_{i}^{\mu}(\theta)},
\end{multline*}
}{
\begin{equation*}
{\mathcal{L}}(\text{data}|\mu,\theta) = \sum_i\frac{\left[\mu s_i(\theta) + b_i(\theta) + \alpha_{\Pe} B_{i}^{\Pe}(\theta) +  \alpha_{\mu} B_{i}^{\mu}(\theta) \right]^{N_i}}{N_i !} \re^{-\mu s_i(\theta) - b_i(\theta) - \alpha_{\Pe} B_{i}^{\Pe}(\theta) - \alpha_{\mu} B_{i}^{\mu}(\theta)},
\end{equation*}
}
\end{linenomath}
where $N_i$ is the observed number of events in each bin, and  $s_i(\theta)$ and $b_i(\theta)$ are the expected signal and background yields in each bin,
respectively, normalized as discussed in the previous sections and taking into account all systematic uncertainties, represented by $\theta$,  as nuisance parameters associated with log-normal priors.
The $B_{i}^{{\Pe}, \mu}(\theta)$ are the yields of NPL backgrounds, and the parameters $\alpha_{{\Pe}, \mu}$, which determine the normalization of the NPL backgrounds, are left free in the fit. The simultaneous fit to the data templates (BDT discriminators or \mtw, depending on the region)  in the four channels maximizes ${\mathcal{L}}(\text{data}|\mu,\theta)$, from which the measured cross section $\sigma(\tZllq)$  is extracted according to its relation to the signal strength
\begin{linenomath}
\begin{equation*}
\mu = \frac{\sigma(\tZllq)}{\sigma^\mathrm{SM}(\tZllq)},
\end{equation*}
\end{linenomath}
where the cross section is defined for any decay of the top quark, and any decay of the \PZ~boson to charged leptons.
The reference cross section is $\sigma^\mathrm{SM}(\tZllq)=94.2$\unit{fb},
for $\Mell >30$\GeV.
The measurement implies an extrapolation from the considered phase space (Section~\ref{sec:CR}), defined as containing
three leptons in the final state
($l'^{+}l'^{-}l$), and an additional constraint for $\Mell$ to be within 15\GeV of the
Z boson mass.
The acceptance, defined as the fraction of \tZllq~events fulfilling the event selection criteria, is estimated from the simulated \tZq~sample as
1.81\%, combining the \tZqCR, \ttZCR,~and \WZCR~regions.
All nuisance parameters are constrained in the fit.

\begin{figure*}
\includegraphics[width=\textwidth]{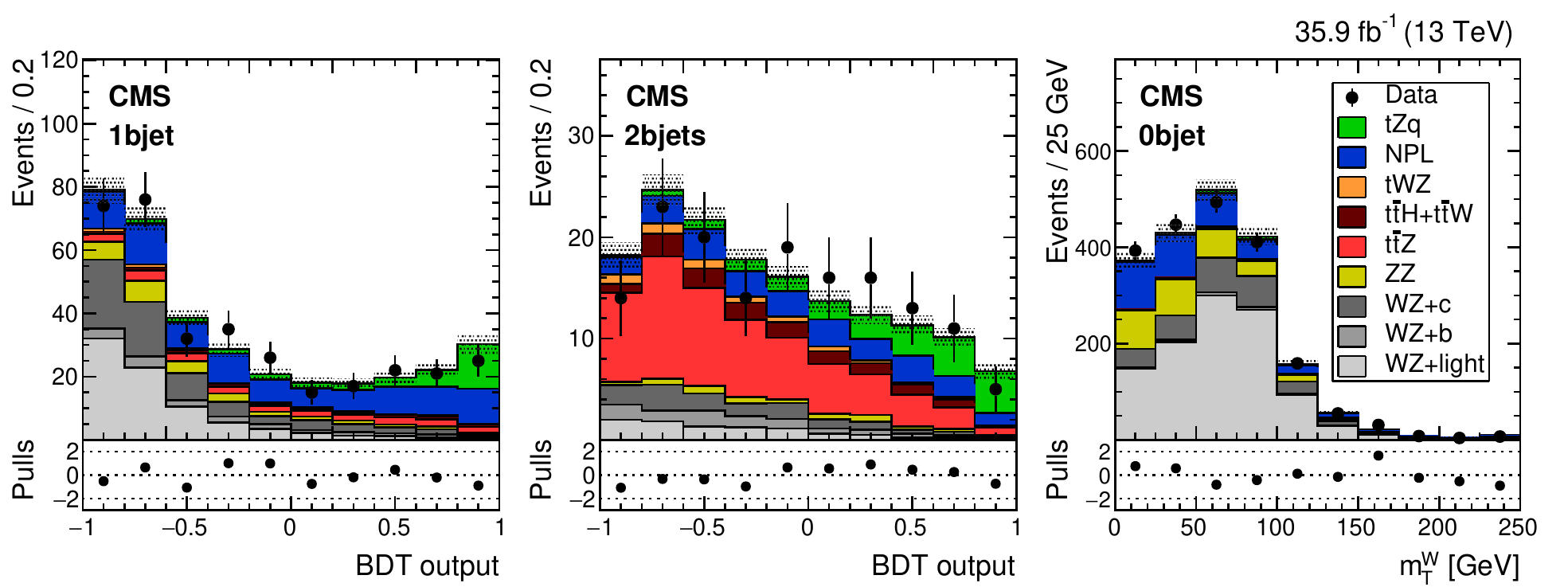}
    \caption{Template distributions used for signal extraction. Left:  BDT discriminator in the \tZqCR~region;
centre: BDT output in the \ttZCR~control region; right: \mtw\ in the \WZCR~control region.
More details are given in the caption of Fig. \ref{fig:signal}.
}
    \label{fig:data-MC-WZ-ptQ}
\end{figure*}

The  distributions resulting from the fit (post-fit)  of the three variables used as templates in the measurement are shown in Fig.~\ref{fig:data-MC-WZ-ptQ}.
Although the fit is performed for each channel, the figure displays the results combining the four channels.

Table \ref{table:signalyields} shows the results for the post-fit yields, separately for each channel, in the \tZqCR~region.
The last two rows show the total number of predicted (``Total") and observed (``Data") events.
The last column displays the ratio of the post-fit to pre-fit predictions, $N^\text{post-fit}/N^\text{pre-fit}$,
accounting for the  systematic uncertainties.
The post-fit background normalizations are
close to the pre-fit values for most of the background processes.
 The event yields for the $\PW\PZ$+light-flavour jets background
preferred by the fit is significantly lower than the SM prediction.
This feature, which might reflect the somewhat worse agreement between simulation and data
 for some bins of jet multiplicity \cite{SMP-16-002},  does
not affect the  measurement, as verified by the following checks. First, the  predicted shapes
of the kinematic variables relevant to the analysis
are verified to describe the data in the \PW\PZ\!+light-flavour enriched region. The analysis
is then  repeated  with the \PW\PZ\!+light-flavour normalization relative
uncertainty increased to 50\%, leaving the results unchanged within about half a percent.
Finally, the \PW\PZ\!+light-flavour
yield is fitted simultaneously with the NPL background yields using only the \WZCR~region, and the resulting
$N^\text{post-fit}/N^\text{pre-fit}$ scale factor is found to be $0.73\pm 0.11$, in good agreement with the results of Table~\ref{table:signalyields}.
The post-fit number of \tZq events in the  \tZqCR~region is 32.3.
The  \WZCR~and \ttZCR~control regions (not shown) also contain \tZq events, with post-fit yields
of $\approx$23 and 19 events,
respectively.

\begin{table*}[hb!t]
\centering
\topcaption{Observed and post-fit expected yields for each production process in the \tZqCR~region. The yields of columns 2--5
correspond to each channel, and  column 6 displays the total for all channels. The last column displays
the ratio between post-fit and pre-fit yields.}
\label{table:signalyields}
\newcolumntype{z}{D{,}{\,\pm\,}{3.3}}
\resizebox{\textwidth}{!}{
 \begin{tabular}{l*{6}{z}}
 Process & \multicolumn{1}{c}{\eee} & \multicolumn{1}{c}{\eemu} & \multicolumn{1}{c}{\mumue} & \multicolumn{1}{c}{\mumumu} & \multicolumn{1}{c}{All channels} & \multicolumn{1}{c}{$N^\text{post-fit}/N^\text{pre-fit}$} \\ \hline
  \tZq & 5.0,1.5 & 6.6,1.9 & 8.5,2.5 & 12.3,3.6 &  32.3,5.0  &  \NA \\
 \ttZ & 3.7,0.7 & 4.7,0.9 & 6.1,1.2 & 8.0,1.5 &  22.4,2.2  &  0.9,0.2 \\
 \ttW & 0.3,0.1 & 0.3,0.1 & 0.7,0.2 & 0.6,0.2 &  1.9,0.3  &  1.0,0.2 \\
 \PZ\!\PZ & 4.8,1.3 & 3.2,0.9 & 9.0,2.5 & 7.8,2.2 &  24.7,3.6  &  1.3,0.3 \\
 \PW\PZ\!+b & 3.0,0.9 & 3.4,1.1 & 4.6,1.4 & 5.5,1.7 &  16.6,2.6  &  1.0,0.2 \\
 \PW\PZ\!+c & 9.0,2.4 & 13.7,3.7 & 18.0,4.9 & 24.2,6.5 &  64.8,9.3  &  1.0,0.2 \\
 \PW\PZ\!+light & 12.2,1.6 & 16.6,2.0 & 22.4,2.8 & 29.1,3.4 &  80.3,5.1  &  0.7,0.1 \\
 \ttH & 0.6,0.2 & 0.9,0.3 & 1.0,0.3 & 1.5,0.4 &  4.0,0.6  &  1.0,0.2 \\
 tWZ & 1.0,0.3 & 1.3,0.4 & 1.7,0.5 & 2.4,0.7 &  6.5,1.0  &  1.0,0.2 \\
 NPL: electrons & 19.2,3.1 & 0.6,0.1 & 17.9,2.8 & \NA &  37.7,4.2  &  \NA \\
 NPL: muons & \NA & 7.2,2.3 & 31.1,9.9 & 15.3,4.9 &  53.6,11.3  &  \NA \\[2ex]
 Total & 58.8,4.8 & 58.4,5.5 & 121,12 & 107,10 &  345,18 &  \\[2ex]
 Data  & 56           & 58          & 104            &  125          &  343            & \\
\end{tabular}
}
\end{table*}

The observed \tZq~signal strength is
\begin{linenomath}
\begin{equation*}
\mu = 1.31^{+0.35}_{-0.33}\stat ^{+0.31}_{-0.25}\syst,
\end{equation*}
\end{linenomath}
from which, using the reference NLO cross section,
 the measured cross section is found to be
\begin{linenomath}
\begin{equation*}
\sigma( \PQt\ell^+\ell^- \PQq)~=123^{+33}_{-31}\stat ^{+29}_{-23}\syst\unit{fb},
\end{equation*}
\end{linenomath}
 for $\Mell>30$\GeV, where $\ell$ stands for electrons, muons, and $\tau$ leptons. The best-fit
signal strength and cross section, as well as an approximate 68\% CL interval, are extracted following the profile likelihood scan procedure described in Ref. \cite{Khachatryan2015}.
The fit is redone without including the systematic uncertainties,
to evaluate the statistical uncertainty of the result. The quoted systematic uncertainty is then calculated as the difference in quadrature
between the  68\% CL intervals obtained in the nominal fit and in the fit without systematic uncertainties.
The precision of the measurement is limited by the statistical uncertainty.
Among the systematic uncertainties, the dominating ones arise from
the normalization of the NPL background (left free in the fit),
the scale dependence at the parton shower level,
the b tagging efficiency, and
the normalization of the \ttZ~background.
The corresponding observed (expected) significance against the background-only hypothesis is 3.7\,(3.1) standard deviations,
with an observed statistical $p$-value of 0.0001. The expected significance is estimated from an Asimov toy dataset \cite{Cowan2011}. The 68\% CL interval of the expected significance is 1.4--5.9.

Potential biases  from the background yields used as input have been searched for.
First, the analysis was repeated to measure simultaneously the \tZq~and \ttZ~cross sections, in addition
to determining  the NPL background normalization. The \tZq~signal strength increases by less
than 1\%, whereas the observed and expected significances decrease by about 1\%.
Then, the not-prompt muon and electron normalizations  are set to their input values, described as the first
step in Section~\ref{sec:NPL}, and allowed to vary in the
fit as Gaussian constraints of 100\% uncertainty. In this case, both the \tZq~signal strength and the significances
 increase by about 10\%, while the uncertainties on the signal strength increase by about 5\%.
In addition, the measurement is repeated in each channel,
and the measured signal strengths are
found to be $1.32^{+1.14}_{-0.99},\ 0.66^{+0.78}_{-0.63},\ 0.01^{+0.97}_{-0.01}$, and $1.22^{+0.75}_{-0.63}$
for the \eee, \eemu, \mumue, and \mumumu~channels, respectively.
The highest observed (expected) significance is 2.1 (1.9) standard deviations in the  \mumumu~channel.
Finally, the results were verified in a counting analysis, using the yields observed in control regions selected
using similar criteria to those of the \tZqCR, \ttZCR, and \WZCR~regions. The simulated backgrounds are
normalized according to their SM predictions, while the normalization of the NPL contributions follows
the procedure described in Section \ref{sec:NPL} as the first step of the NPL normalization in the shape analysis.
The results from the counting and shape analyses are in agreement.

\section{Summary}

The associated production cross section of a single top quark and a \PZ boson was measured using
data from $\Pp\Pp$~collisions at 13\TeV collected by the CMS experiment, corresponding to
an integrated luminosity of 35.9\fbinv.
The measurement uses events containing three charged leptons in the final state.
Evidence for \tZq~production is found with an observed (expected) significance of 3.7\,(3.1) standard deviations.
The cross section is
measured to be $\sigma( \tZllq)= 123^{+33}_{-31}\stat ^{+29}_{-23}\syst\unit{fb}$,
for $\Mell>30$\GeV, where
$\ell$ stands for electrons, muons, or $\tau$ leptons.
This value  is compatible with the next-to-leading-order standard model prediction of $94.2\pm3.1$\unit{fb}.

\begin{acknowledgments}
We congratulate our colleagues in the CERN accelerator departments for the excellent performance of the LHC and thank the technical and administrative staffs at CERN and at other CMS institutes for their contributions to the success of the CMS effort. In addition, we gratefully acknowledge the computing centres and personnel of the Worldwide LHC Computing Grid for delivering so effectively the computing infrastructure essential to our analyses. Finally, we acknowledge the enduring support for the construction and operation of the LHC and the CMS detector provided by the following funding agencies: BMWFW and FWF (Austria); FNRS and FWO (Belgium); CNPq, CAPES, FAPERJ, and FAPESP (Brazil); MES (Bulgaria); CERN; CAS, MoST, and NSFC (China); COLCIENCIAS (Colombia); MSES and CSF (Croatia); RPF (Cyprus); SENESCYT (Ecuador); MoER, ERC IUT, and ERDF (Estonia); Academy of Finland, MEC, and HIP (Finland); CEA and CNRS/IN2P3 (France); BMBF, DFG, and HGF (Germany); GSRT (Greece); OTKA and NIH (Hungary); DAE and DST (India); IPM (Iran); SFI (Ireland); INFN (Italy); MSIP and NRF (Republic of Korea); LAS (Lithuania); MOE and UM (Malaysia); BUAP, CINVESTAV, CONACYT, LNS, SEP, and UASLP-FAI (Mexico); MBIE (New Zealand); PAEC (Pakistan); MSHE and NSC (Poland); FCT (Portugal); JINR (Dubna); MON, RosAtom, RAS, RFBR and RAEP (Russia); MESTD (Serbia); SEIDI, CPAN, PCTI and FEDER (Spain); Swiss Funding Agencies (Switzerland); MST (Taipei); ThEPCenter, IPST, STAR, and NSTDA (Thailand); TUBITAK and TAEK (Turkey); NASU and SFFR (Ukraine); STFC (United Kingdom); DOE and NSF (USA).

\hyphenation{Rachada-pisek} Individuals have received support from the Marie-Curie programme and the European Research Council and Horizon 2020 Grant, contract No. 675440 (European Union); the Leventis Foundation; the A. P. Sloan Foundation; the Alexander von Humboldt Foundation; the Belgian Federal Science Policy Office; the Fonds pour la Formation \`a la Recherche dans l'Industrie et dans l'Agriculture (FRIA-Belgium); the Agentschap voor Innovatie door Wetenschap en Technologie (IWT-Belgium); the Ministry of Education, Youth and Sports (MEYS) of the Czech Republic; the Council of Science and Industrial Research, India; the HOMING PLUS programme of the Foundation for Polish Science, cofinanced from European Union, Regional Development Fund, the Mobility Plus programme of the Ministry of Science and Higher Education, the National Science Center (Poland), contracts Harmonia 2014/14/M/ST2/00428, Opus 2014/13/B/ST2/02543, 2014/15/B/ST2/03998, and 2015/19/B/ST2/02861, Sonata-bis 2012/07/E/ST2/01406; the National Priorities Research Program by Qatar National Research Fund; the Programa Severo Ochoa del Principado de Asturias; the Thalis and Aristeia programmes cofinanced by EU-ESF and the Greek NSRF; the Rachadapisek Sompot Fund for Postdoctoral Fellowship, Chulalongkorn University and the Chulalongkorn Academic into Its 2nd Century Project Advancement Project (Thailand); the Welch Foundation, contract C-1845; and the Weston Havens Foundation (USA).
\end{acknowledgments}

\bibliography{auto_generated}
\clearpage
\appendix

\section{Variables used in the shape analysis}
\label{appBDT}

A short description of the variables used in the Boosted Decision Trees in the \tZqCR~and \ttZCR~regions
 is given in Table~\ref{table:BDTvar}. The BDTs in the \tZqCR~region
and \ttZCR~region used 16 and 15 variables each, respectively, 10 of which common to the two regions
(variables 1--9 and 17 in Table~\ref{table:BDTvar}).
From the total of 21 variables, 16 are related to the kinematic quantities associated to the final state objects
(1--16 in Table~\ref{table:BDTvar}),
while five of them are related to the MEM (17--21 in Table~\ref{table:BDTvar}).

\begin{table*}[hb!t]
\centering
\topcaption{Description of the variables used in the BDTs. The symbol Y (N)
in the third and fourth columns
indicates that the variable was (was not) used in the \tZqCR~and \ttZCR~BDTs.
\label{table:BDTvar}}
\resizebox{\textwidth}{!}{
 \begin{tabular}{llcc}
  & Variable description & \tZqCR & \ttZCR  \\ \hline
1 & CSVv2 algorithm discriminant & Y & Y \\
2 & $\Delta R$ separation between the jet identified as a b quark and the recoiling jet & Y & Y \\
3 & $\eta$ of the recoiling jet  & Y & Y \\
4 & \pt of the recoiling jet  & Y & Y \\
5 &  $\eta$ of the \PZ~boson  & Y & Y \\
6 & Top quark mass  & Y & Y \\
7 &  $\Delta R$ separation between the top quark decay lepton and the jet closest to it  & Y & Y \\
8 & Top quark decay lepton asymmetry & Y & Y \\
9 & Azimuth angle separation between the top quark decay lepton and the \PZ~boson  & Y & Y \\
10 & Azimuth angle separation between the top quark decay lepton and the b quark  & Y & N \\
11 & $\eta$ of  the top quark decay lepton  & Y & N \\
12 & $\eta$ of the jet with highest \pt & Y & N \\
13 & $\Delta R$ separation between the top quark decay lepton and the recoil jet  & N & Y \\
14 & $\Delta R$ separation between the \Z~boson and the top quark  & N & Y \\
15 & \pt of the  \Z~boson  & N & Y \\
16 & Number of b tagged jets  & N & Y \\
17 & Logarithm of the MEM score associated to the most probable \tZq~kinematic configuration  & Y & Y \\
18 & Logarithm of the MEM score associated to the most probable \ttZ~kinematic configuration  & N & Y \\
19 & Log-likelihood ratio of the \tZq~hypothesis against the \ttZ~hypothesis  & Y & N \\
20 & Log-likelihood ratio of the \tZq~hypothesis against the \ttZ~hypothesis  & Y & N \\
 & with \ttZ~and \tZq~weights rescaled such that their mean values are similar & & \\
21 & Log-likelihood ratio of the MEM weights for \ttZ~against \ttZ+\PW\PZ~hypothesis & Y & N \\
\end{tabular}
}
\end{table*}
\cleardoublepage \section{The CMS Collaboration \label{app:collab}}\begin{sloppypar}\hyphenpenalty=5000\widowpenalty=500\clubpenalty=5000\input{TOP-16-020-authorlist.tex}\end{sloppypar}
\end{document}

%% file: TOP-16-020-authorlist.tex
\textbf{Yerevan Physics Institute,  Yerevan,  Armenia}\\*[0pt]
A.M.~Sirunyan, A.~Tumasyan
\vskip\cmsinstskip
\textbf{Institut f\"{u}r Hochenergiephysik,  Wien,  Austria}\\*[0pt]
W.~Adam, F.~Ambrogi, E.~Asilar, T.~Bergauer, J.~Brandstetter, E.~Brondolin, M.~Dragicevic, J.~Er\"{o}, A.~Escalante Del Valle, M.~Flechl, M.~Friedl, R.~Fr\"{u}hwirth\cmsAuthorMark{1}, V.M.~Ghete, J.~Grossmann, J.~Hrubec, M.~Jeitler\cmsAuthorMark{1}, A.~K\"{o}nig, N.~Krammer, I.~Kr\"{a}tschmer, D.~Liko, T.~Madlener, I.~Mikulec, E.~Pree, N.~Rad, H.~Rohringer, J.~Schieck\cmsAuthorMark{1}, R.~Sch\"{o}fbeck, M.~Spanring, D.~Spitzbart, A.~Taurok, W.~Waltenberger, J.~Wittmann, C.-E.~Wulz\cmsAuthorMark{1}, M.~Zarucki
% ...
\vskip\cmsinstskip
% ...
1:~~Also at Vienna University of Technology, Vienna, Austria\\
% ...
